\documentclass[12pt]{article}
\usepackage[utf8]{inputenc}
\usepackage{amssymb, amsmath}
\usepackage{appendix}
\usepackage{setspace}
\usepackage{graphicx}
\usepackage{morefloats}
\usepackage[numbers]{natbib}
\usepackage{hyperref}
\usepackage{multirow}
\usepackage{slashed}
\usepackage{mathrsfs}

\newcommand{\ud}{\mathrm{d}}

\hypersetup{
  colorlinks = true,
  urlcolor = black,
  linkcolor= blue,
  citecolor= green,
  filecolor= magenta,
}

\title{Bounds on Large Extra Dimensions from the Generalized Uncertainty Principle}
\author{
Marco Cavagli\`a\thanks{Email: cavaglia$@$phy.olemiss.edu}
 \\ \small{Department of Physics \& Astronomy, The University of Mississippi}
 \\ \small{University, Mississippi 38677-1848, USA}
\and
Benjamin Harms\thanks{Email: bharms$@$ua.edu},
Shaoqi Hou\thanks{Email: shou$@$crimson.ua.edu}
\\ \small{Department of Physics \& Astronomy, The University of Alabama}
\\ \small{Tuscaloosa, Alabama 35487-0324, USA}
}
\date{\today}

\begin{document}
\maketitle

\abstract{
The Generalized Uncertainty Principle (GUP) implies the existence of a physical minimum length scale $l_m$. In this scenario, black holes must have a radius larger than $l_m$. They are hotter and evaporate faster than in standard Hawking thermodynamics. We study the effects of the GUP on black hole production and decay at the LHC in models with large extra dimensions. Lower bounds on the fundamental Planck scale and the minimum black hole mass at formation are determined from black hole production cross section limits by the CMS Collaboration. The existence of a minimum length generally decreases the lower bounds on the fundamental Planck scale obtained in the absence of a minimum length.
}

\section{Introduction}

Models of Large Extra Dimensions (LEDs) \cite{aad98,add98,add99} open the possibility that the fundamental Planck scale may be at the TeV energy scale. This has stimulated interest in studying black hole (BH) production at the Large Hadron Collider (LHC) and future particle accelerators. Many quantum gravity candidates, such as, string theory \cite{AMATI198941,AMATI1990550,AMATI1993707,Konishi1990276}, noncommutative geometry \cite{Maggiore:1993zu,Maggiore:1993kv} and loop quantum gravity \cite{Rovelli:1994ge}, seem to suggest a modification of the Heisenberg uncertainty principle at or around the Planck scale. The modification has become known as the Generalized Uncertainty Principle (GUP). The GUP implies the existence of a minimum length scale $l_m$ of the order of the Planck length $l_\mathrm{Pl}$. If that is the case, BHs produced at the LHC would have a radius larger than $l_m$. The GUP also affects the subsequent decay of BHs via Hawking radiation \cite{hawkeff} by making them hotter, shorter-lived, and reducing their entropy. It is thus interesting to investigate in detail how the GUP may affect the phenomenology of BH production and decay at the LHC.

According to the Hoop Conjecture \cite{kt}, a BH forms when a mass $M$ is confined to a region of size comparable to the Schwarzschild radius for that mass  \cite{mpDbh,mc,kantiReview},
\begin{equation}\label{rs}\displaystyle
R_\mathrm{S}=\frac{1}{\sqrt{\pi}M_*}\left[\frac{8\Gamma\left(\frac{n+3}{2}\right)}{n+2}\right]^{\frac{1}{n+1}}\left(\frac{M}{M_*}\right)^{\frac{1}{n+1}},
\end{equation}
where $n$ is the number of LEDs and $M_*$ is the fundamental Planck scale.
Therefore, if two particles collide with center of mass energy $\sqrt{s}$ and impact parameter smaller than $R_\mathrm{S}(\sqrt{s})$, a BH may form.
The Hoop Conjecture suggests that the BH production cross section is of the same order as that of the black disk cross section $\sigma(\sqrt s,n)\sim\pi R^2_\mathrm{S}$. Since BH production of hadron colliders occurs at the parton level, the total cross section for BH production in a hadronic collision is obtained by integrating over the Parton Distribution Functions (PDFs) of the hadrons \cite{mc}
\begin{equation}\label{sigmatt}
\sigma(s,n)=\sum_{ij}\int_0^12z\ud z\int_{x_m}^1\ud x\int_x^1\frac{\ud x\rq{}}{x\rq{}}f_i(x\rq{},Q)f_j(x/x\rq{},Q)\sigma(\sqrt{xs},n),
\end{equation}
where $f_i(x,Q)$ are the PDFs with four-momentum transfer $Q$ and $z$ is the impact parameter normalized to its maximum value. The cutoff at small $x$ is $x_m={M_\mathrm{min}}^2/s$, where $M_\mathrm{min}$ is the minimum-allowed mass of the BH. If the initial BH mass is much larger than the Planck mass, a semiclassical treatment suggests that the newly-formed BH decays through four, possibly distinct stages \cite{kantiReview}: a \emph{balding} phase, where the BH radiates multipole momenta and quantum numbers \cite{giddings, gl} eventually settling down to a $D$-dimensional Kerr geometry;  a \emph{spin-down} phase \cite{gl}, where angular momentum is radiated;  a \emph{Hawking} phase, where the Schwarzschild BH decays into elementary particles through the Hawking mechanism;  a \emph{Planck} phase, where the mass of the BH approaches the Planck scale and the decay becomes dominated by quantum gravitational effects.

The Compact Muon Solenoid (CMS) Collaboration has conducted several searches for BH events \cite{Khachatryan:2010wx,CMS:2012yf,Chatrchyan:2012taa,exo-12-009,Khachatryan:2015sja}. In one of their most recent publications,  model-independent upper cross section limits for BH production were obtained by analyzing a data sample of proton-proton ($pp$) collisions at $\sqrt{s}=8$ TeV and integrated luminosity of 12.1 fb$^{-1}$ \cite{exo-12-009}. These cross section limits can be used to set bounds on $M_*$ and the minimum BH mass $M_\mathrm{min}$ at formation. The CMS Collaboration used the reduced Planck scale $M_D$ to characterize the size of the LEDs. $M_D$ is related to $M_*$ by
\begin{equation}\label{eq}
  M_D=\Big[\frac{(2\pi)^n}{8\pi}\Big]^{\frac{1}{n+2}}M_*.
\end{equation}
In this work, we use the CMS model-independent results to determine these bounds in the presence of the GUP. In the next section, we present a brief discussion on the GUP and its implications for BH formation and decay. Section \ref{sec-catfish} is devoted to the description of the BH event generator CATFISH \cite{catfish}. Results and conclusions are presented in Section \ref{sec-results} and Section \ref{sec-final}, respectively.

\section{Generalized Uncertainty Principle}\label{sec-gup}

The GUP is closely related to the postulate of a minimum length scale $l_m$, dating back to 1947, when Synder \cite{PhysRev.71.38} proposed that spacetime may be discrete if spacetime coordinates are noncommutative. Later, Mead \cite{PhysRev.135.B849,PhysRev.143.990} showed that it is impossible to measure distances less than the Planck length and Majid and Ruegg \cite{MAJID1994348} modified the Poincar\'e algebra to include nonvanishing commutators between spacetime coordinates (constituting $\kappa$-Poincar\'e algebra) resulting in a minimum length scale. Investigations with noncommutative geometry \cite{Maggiore:1993zu,Maggiore:1993kv}, string theory \cite{AMATI198941,AMATI1990550,AMATI1993707,Konishi1990276} and loop quantum gravity \cite{Rovelli:1994ge} also predict a minimum length scale. For more history on this subject, please refer to Ref.\cite{Hossenfelder:2012jw}.

In quantum mechanics, the Heisenberg uncertainty principle $\Delta x \Delta p\gtrsim \hbar$ implies that more and more energy is required to probe smaller and
smaller distances. As long as the energy available is large enough, an
arbitrarily small distance can be probed. If a physical minimum length
exists, the Heisenberg uncertainty principle must be modified at high
energies. Assuming that the corrections to the Heinsenberg principle become
relevant at the Planck scale, the most common version of a GUP in $D=n+4$
dimensions can be written as \cite{Cavaglia:2003qk,Cavaglia:2004jw},
\begin{equation}\label{gup}
\Delta x_i\gtrsim \frac{\hbar}{\Delta p_i}+\alpha^2l_\mathrm{Pl}^2\frac{\Delta p_i}{\hbar},
\end{equation}
where $l_\mathrm{Pl}=(\hbar G_D/c^3)^{1/(D-2)}$ is the $D$-dimensional Planck length, $\alpha$ is a dimensionless constant of order one, and the index $i=1,...,n+3$ labels spatial coordinates. Equation (\ref{gup}) implies the minimum length
\begin{equation}
l_m=2\alpha l_\mathrm{Pl}.
\end{equation}
If the GUP is realized in nature, the diameter of a BH must be at least $l_m$ and its mass must be greater than \cite{Cavaglia:2003qk,Cavaglia:2004jw}
\begin{equation}\label{mmin}
M_\mathrm{ml}=\frac{n+2}{8\Gamma(\frac{n+3}{2})}(\alpha\sqrt{\pi})^{n+1} M_*.
\end{equation}
A BH can form only if its mass $M_\mathrm{BH}$ is larger
than both $M_\mathrm{ml}$ and $M_*$. Since $M_\mathrm{ml}/M_*$ is a monotonically increasing function
of $\alpha$,  the energy above which a BH may form is $M_u=\mathrm{Max}\{M_*,M_\mathrm{ml}\}$. The GUP also leads to modifications of the laws of BH thermodynamics.
The GUP-modified Hawking temperature is
\begin{equation}
T\rq{}_\mathrm{H}=T_\mathrm{H}\frac{2}{1+\sqrt{1-(\alpha l_\mathrm{Pl}/R_\mathrm{S})^2}},
\end{equation}
where $T_\mathrm{H}$ is the standard Hawking temperature
\begin{equation}\label{eq-ht}
T_H=\frac{n+1}{4\pi R_\mathrm{S}}.
\end{equation}
The modified Hawking temperature is higher than the standard Hawking
temperature and monotonically increases with $\alpha$. This implies
that a BH in the GUP scenario radiates at a faster rate than in the standard scenario. The GUP-modified BH
entropy is
\begin{equation}
S\rq{}_H=2\pi\alpha\Big(\frac{\alpha}{M_*R_\mathrm{S}}\Big)^{n+1}\frac{M_\mathrm{BH}}{M_*} I\Big(1, n, \frac{M_*R_\mathrm{S}}{\alpha}\Big),
\end{equation}
where $I(p, q, x)=\int_1^xz^q(z+\sqrt{z^2-1})^pdz$. The modified entropy $S'_\mathrm{H}$ is always less than the standard Hawking entropy
\begin{equation}\label{eq-s0i}
S_H=\frac{n+1}{n+2}\frac{M_\mathrm{BH}}{T_H}
\end{equation}
with the ratio $S'_\mathrm{H}/S_\mathrm{H}$ decreasing monotonically as $\alpha$ increases.
Therefore, the BH decay multiplicity during the Hawking phase in the GUP
scenario is less than in the standard Hawking scenario, the effect
becoming more significant for a larger minimum length. The GUP also leads to the termination of the BH radiation when the BH mass reaches $M_u$ \cite{Cavaglia:2003qk,Cavaglia:2004jw}.

\section{CATFISH}\label{sec-catfish}

In this analysis, we use CATFISH \cite{catfish} to simulate BH production and decay processes. CATFISH is a Fortran 77 Monte Carlo generator that is designed specifically for simulating BH events at CERN's LHC and incorporates the effects of the GUP. The generator interfaces to the PYTHIA Monte Carlo fragmentation code \cite{Sjostrand:2006za,Sjostrand:2014zea} using the Les Houches interface \cite{Boos:2001cv}.
To determine the physics of BH formation and decay CATFISH uses several external parameters and switches: $M_*$, $n$, and the parameter $\alpha$ that determines whether the GUP scenario is turned on ($\alpha>0$) or off ($\alpha=0$). The parameter \texttt{XMIN}$\ge1$ gives the minimum BH mass at the formation, i.e., $\texttt{XMIN}=M_\mathrm{min}/M_*$. $\texttt{QMIN}=Q_\mathrm{min}/ M_*$ determines the BH mass at the onset of the Planck phase. $\texttt{NP} $ is the number of quanta in the final $n$-body decay of the Planck phase. If $\texttt{NP}=0$,  BHs form stable remnants with mass $Q_\mathrm{min}$.
%
%
%
%
%
%
%
%
%
%
  CATFISH is configured to generate parton level events in LHEF format \cite{Alwall2007300}, which are fed into PYTHIA (v8.2.12) to hadronize. 
We run CATFISH (v2.10) with the CTEQ6L1 PDF set and PYTHIA Tune Z1.

Equation (\ref{mmin}) implies that a BH can form at center of mass energy $\sqrt{s}$ only when alpha is smaller than
\begin{equation}\label{eq-predafm}
\alpha_m=\frac{1}{\sqrt{\pi}}\Big[\frac{8\Gamma(\frac{n+3}{2})}{n+2}\frac{\sqrt{s}}{M_*\texttt{XMIN}}\Big]^{\frac{1}{n+1}},
\end{equation}
where $\alpha_m$ is a decreasing function of $M_*$ and \texttt{XMIN}. In the following analysis, the values of $\alpha$ chosen are smaller than $\alpha_m$ so that BHs must form in the simulation.

\section{Lower Bounds on $M_*$ and $M_\mathrm{min}$ from the GUP}\label{sec-results}

The lower bounds on $M_*$ and $M_\mathrm{min}$ (or equivalently, \verb+XMIN+) are obtained using the same method as used in Ref.\cite{Hou:2015gba}. As the results do not significantly depend on $n$ and \verb+NP+,  we  present the bounds only for $n=4, 6$ and $\verb+NP+=0,4$. To see the effects of the minimum length without suppressing BH production, we let $\alpha$ vary between 0.2 and 0.7 so that $M_\mathrm{ml} < M_*$. Additionally, we  run CATFISH with  values of $\alpha=0.9$  and $\alpha=1.0$. These values allow us to estimate the effects of a highly-suppressed cross section.

\begin{figure}[!ht]
\center
\includegraphics[height=6cm]{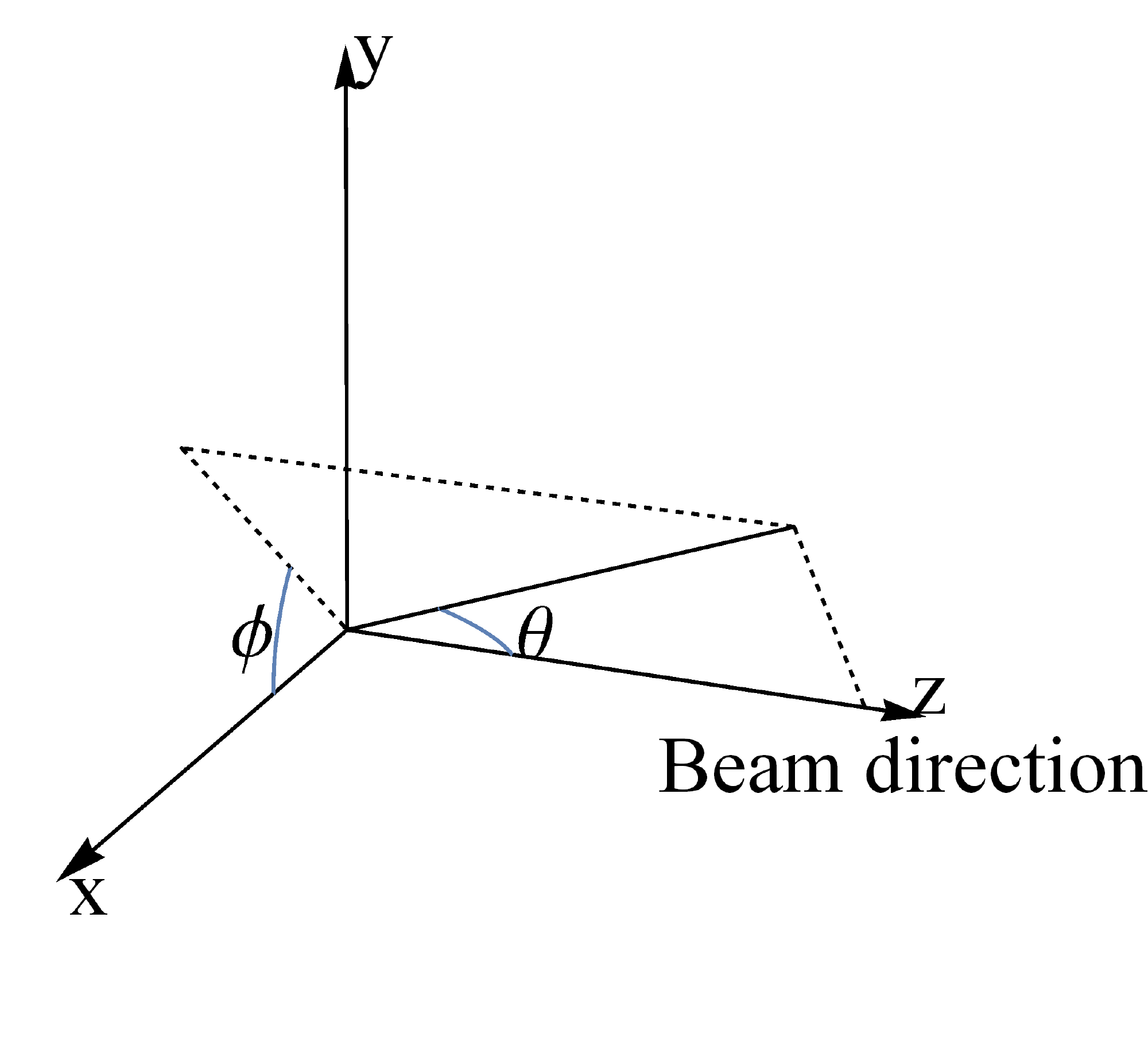}
\includegraphics[height=6cm]{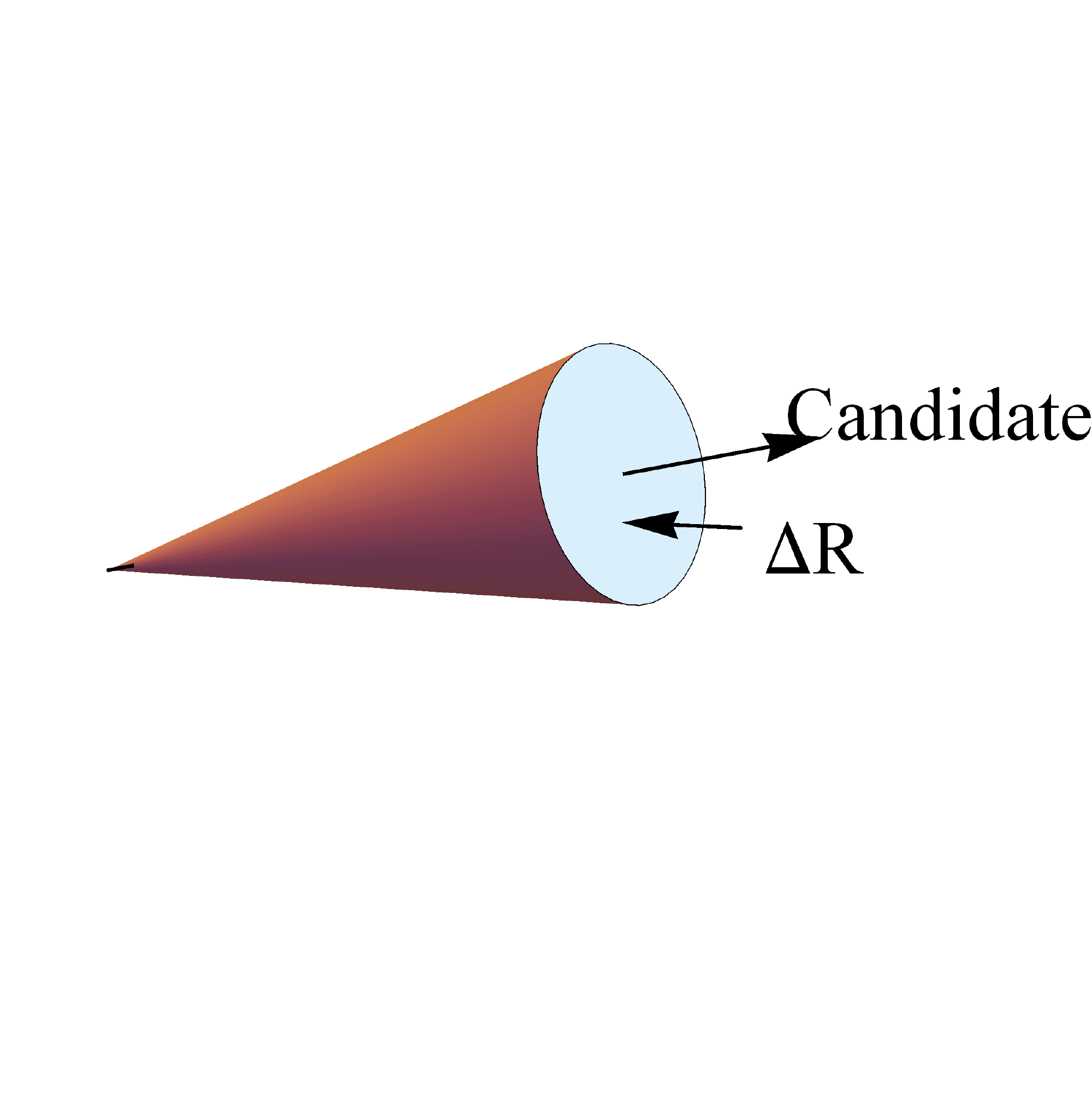}
\caption{\small Left: The coordinate system of the CMS detector. The beam direction is parallel or anti-parallel to the $z$ direction. $y$ is pointing upward and $x$ is pointing into the center of the LHC ring. Right: The cone used to define the variable Iso$(\iota)$. }
\label{fig-cmsc}
\end{figure}

We use Delphes (v3.3.0) \cite{deFavereau:2013fsa} to perform the detector simulation. The FastJet algorithm \cite{Cacciari:2011ma} provides fast naive implementations of different jet-finding algorithms, including the anti-$k_{\rm{T}}$ algorithm used in this analysis. The final-state objects are selected according to the kinematic cuts summarized in Table \ref{tab-kcuta}. The variable Iso($\iota$) refers to the isolation requirement for a particular object (electron, muon or photon). For an electron or a muon candidate, it is the sum of the transverse momenta $p_\mathrm{T}$ of all charged and neutral particles in a cone of $\Delta R=\sqrt{(\Delta \phi)^2+(\Delta\eta)^2}\equiv\iota$ around the object, whereas for a photon candidate it is the ratio of that sum to the $p_\mathrm{T}$ of the photon candidate. Here, $\eta$ is the pseudorapidity of the particle,
\begin{equation}
\eta=\frac{1}{2}\ln\frac{p+p_z}{p-p_z}=-\ln\Big(\tan\frac{\theta}{2}\Big),
\end{equation}
where $\theta$ is the polar angle, and $\phi$ is the azimuthal angle defined in Figure \ref{fig-cmsc}.
\begin{table}[!ht]
\center
\def\arraystretch{1.1}
\begin{tabular}{c*{3}{|c}}
\hline\hline
Objects & $p_\mathrm{T}$ & $|\eta|$ & Iso($\iota$) \\
\hline
Jets &  \multirow{4}{*}{$>50$ GeV} & $<2.6$ & N/A\\
\cline{1-1}\cline{3-4}
Muons & & $<2.1$ & \multirow{2}{*}{Iso$(0.3)<20\%$}\\
\cline{1-1}\cline{3-3}
Electrons & & (1.56, 2.4) \& & \\
\cline{1-1}\cline{4-4}
Photons & &  (0, 1.44) & $\dagger$ \\
\hline
\end{tabular}
\caption{\small Kinematic cuts for the $S_\mathrm{T}$ spectrum. $\dagger$: The scalar sums of transverse energy (momenta in the case of the tracker) of all particles are calculated in a cone of $\Delta R=0.4$ around the candidate photon direction. They are smaller than 2.0, 4.2, and 2.2 GeV for the tracker, ECAL, and HCAL, respectively.}
\label{tab-kcuta}
\end{table}
The minimum separation between any two objects in the event is required to be $\Delta R>0.3$. These requirements match, for the most part, the requirement that are used by the CMS Collaboration in Ref.\cite{exo-12-009}, except for the requirements on the muon impact parameter, the separation between an electron candidate and a muon candidate with more than 10 hits in the inner tracker, and the ratio of HCAL to ECAL energy deposits for a photon candidate. These last three requirements cannot be implemented in Delphes. However, they do not significantly affect Delphes' output \cite{deFavereau:2013fsa}, and can be safely ignored.

\begin{figure}[!ht]
\center
\includegraphics[height=7.1cm]{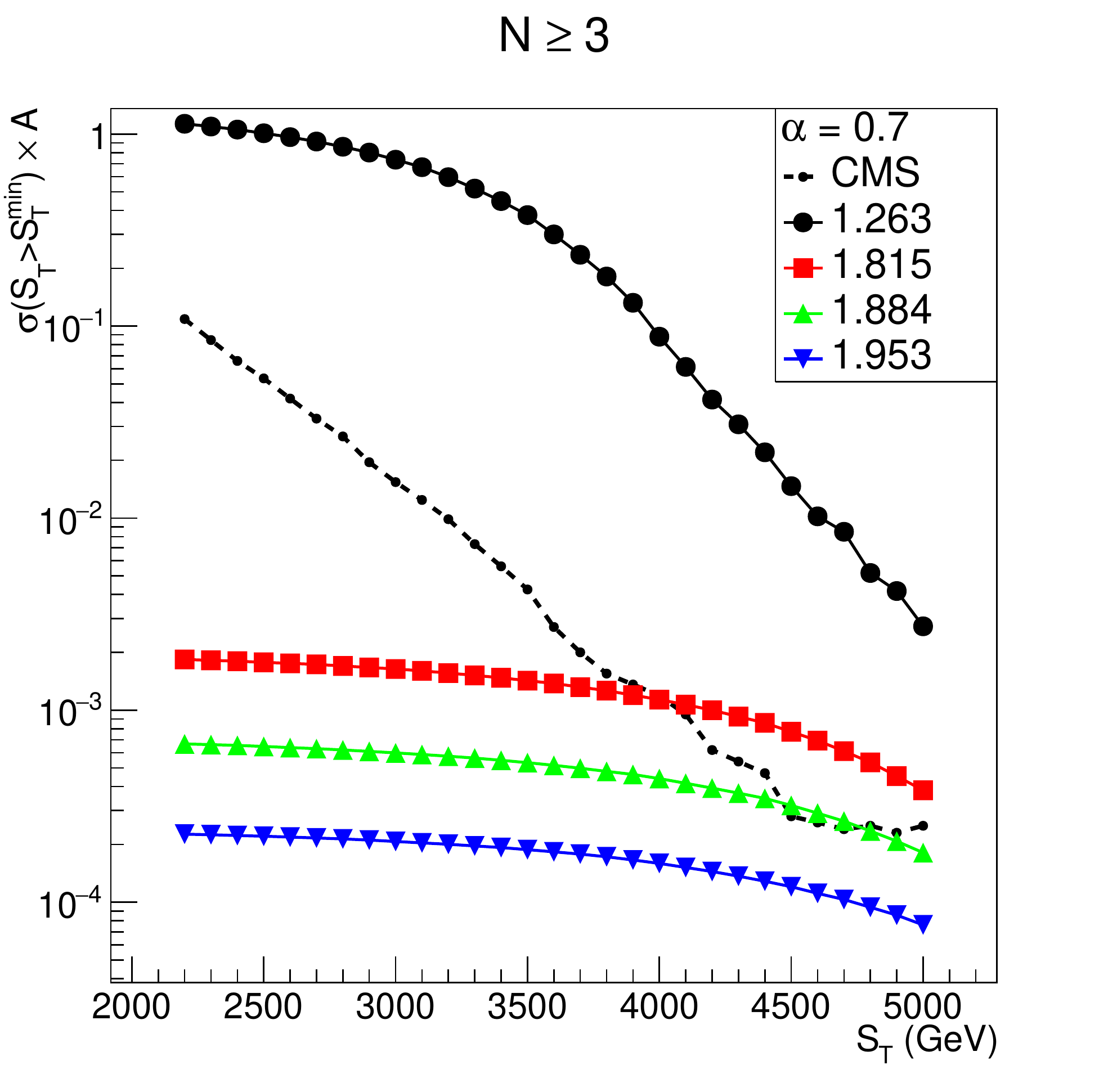}
\includegraphics[height=7.1cm]{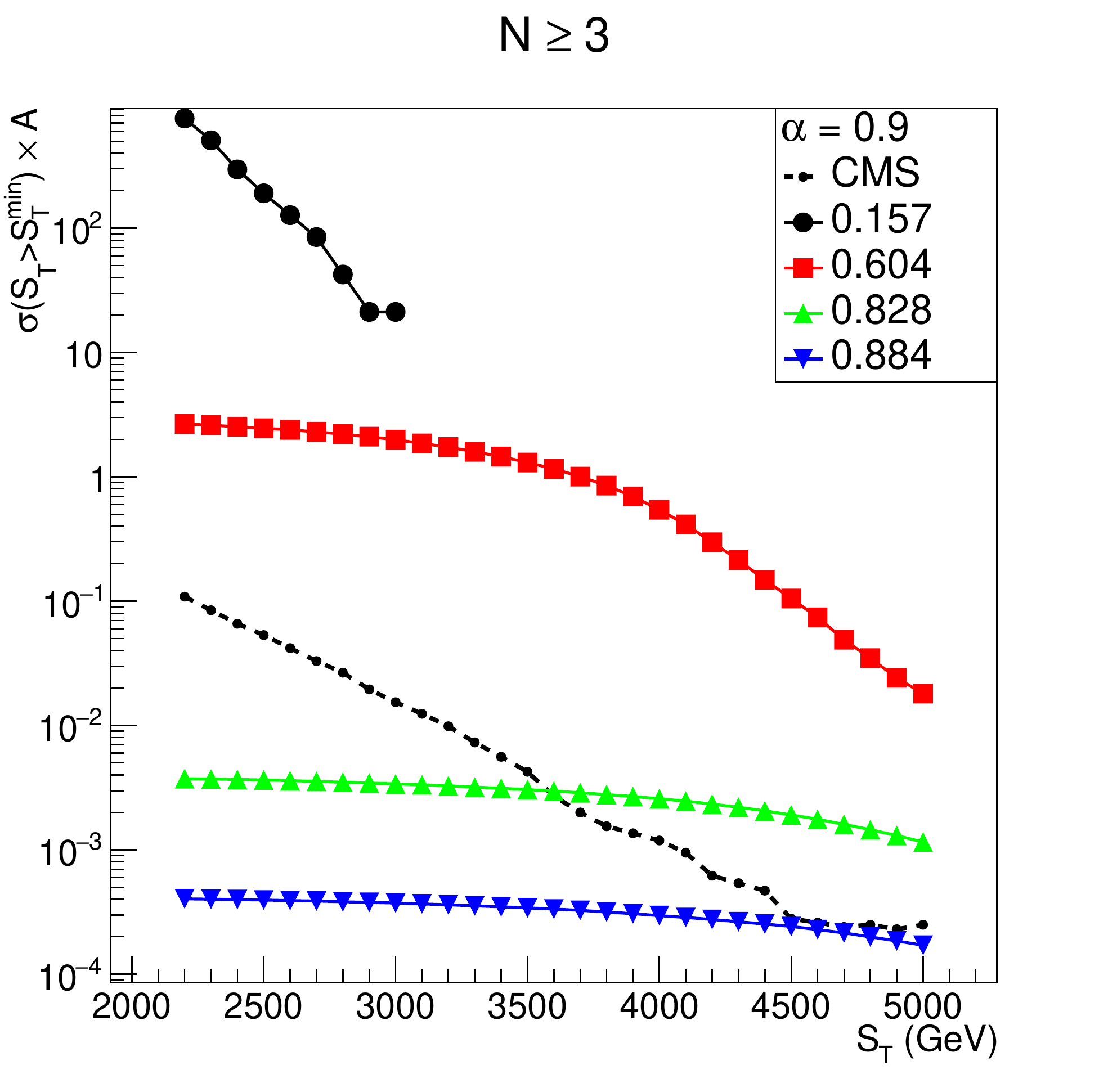}

\includegraphics[height=7.1cm]{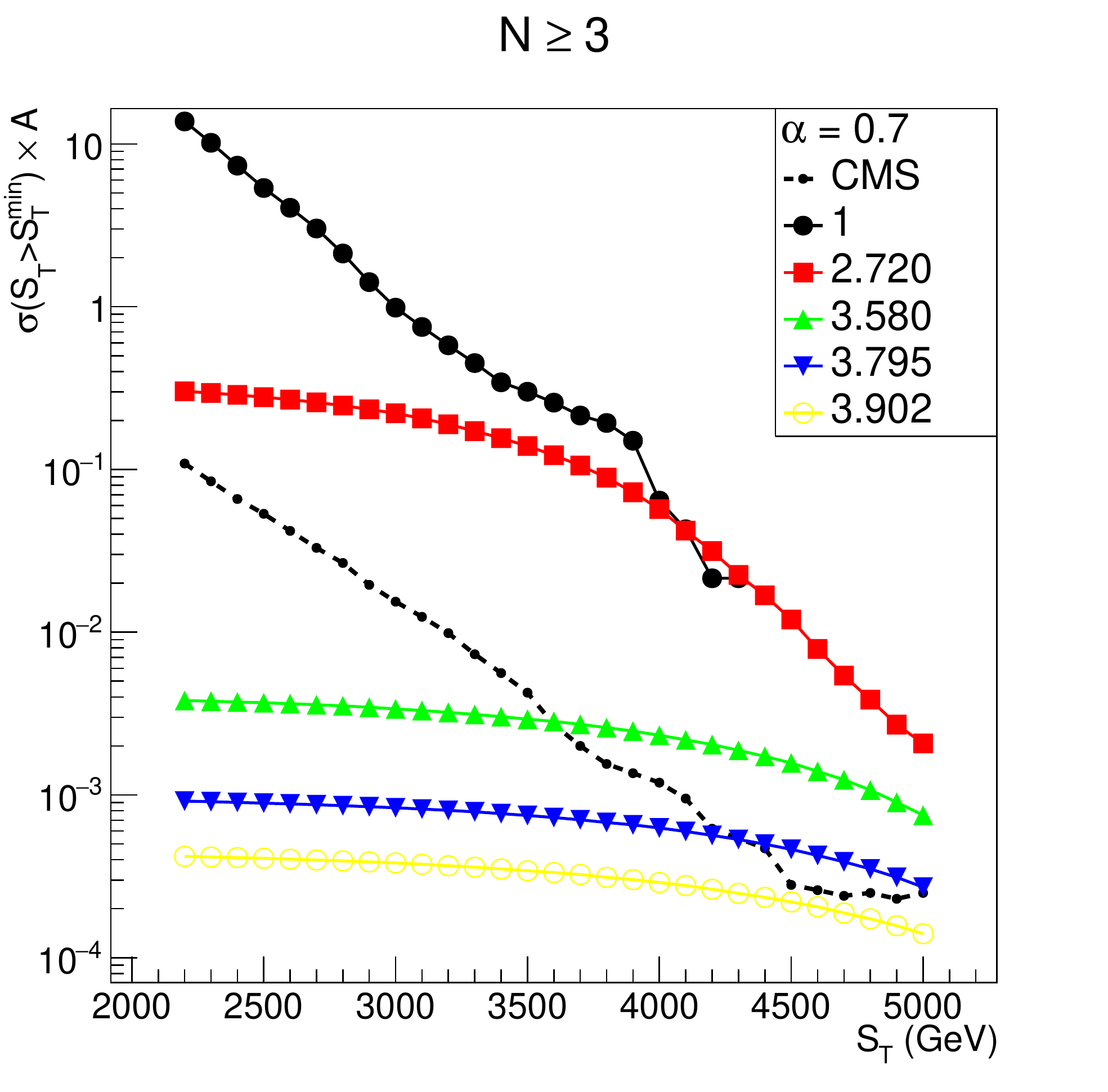}
\includegraphics[height=7.1cm]{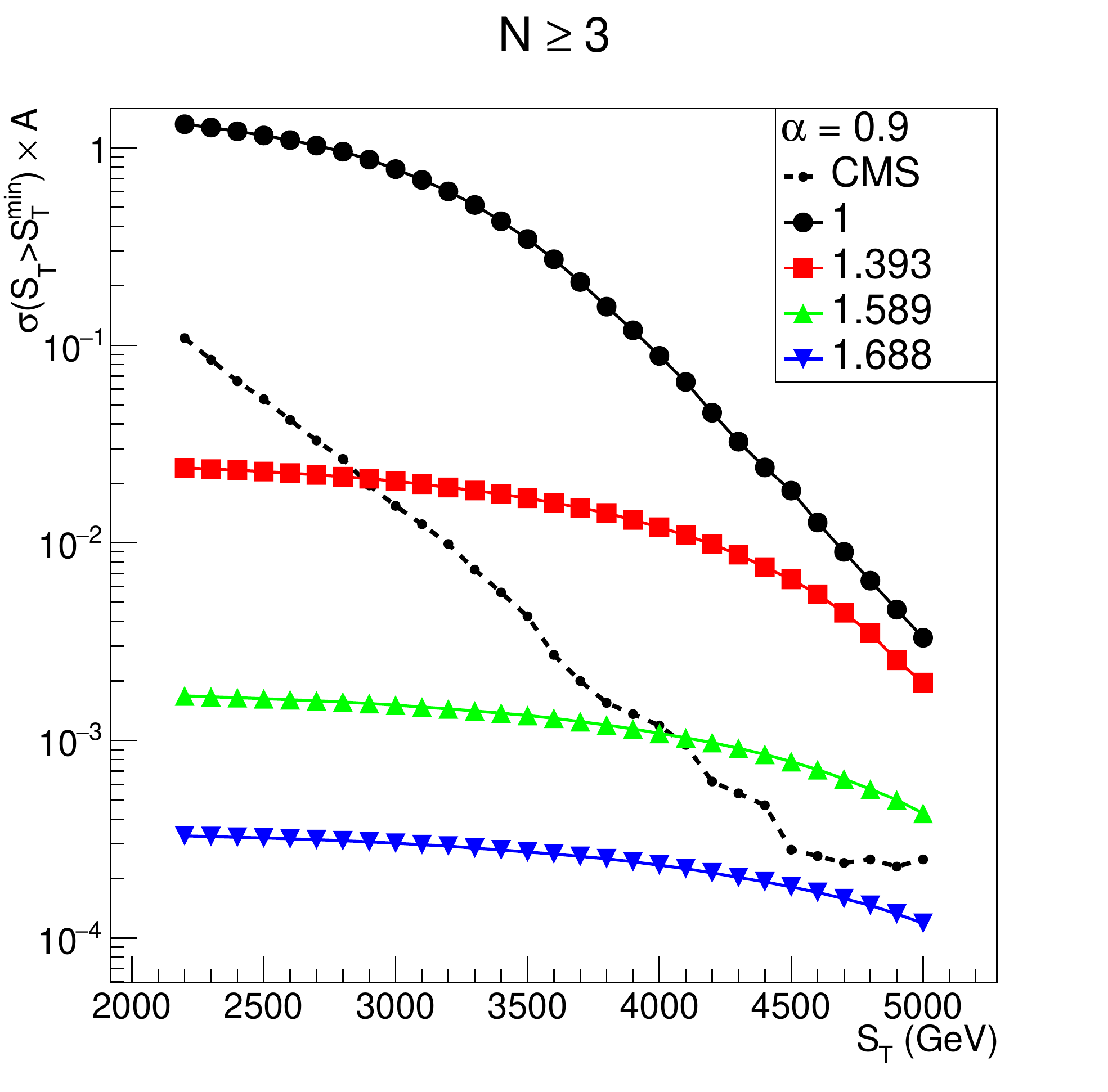}
\caption{\small $\sigma(S_\mathrm{T}>S_\mathrm{T}^\mathrm{min})\times A$ as a function of $M_*$ (upper plots, the values in the legends are for $M_*$ in units of TeV) or \texttt{XMIN} (lower plots, the values in the legend are for \texttt{XMIN}'s) at $\alpha=0.7$ (left plots) and $\alpha=0.9$ (right  plots). The model-independent 95\% CL experimental upper limits from the CMS Collaboration are also shown \cite{exo-12-009}. The event multiplicity is $N\ge 3$.}
\label{fig-xsecAcca}

\end{figure}

Lower bounds on $M_*$ and $M_\mathrm{min}$ are derived by evaluating the partial cross section $\sigma(S_\mathrm{T}>S_\mathrm{T}^\mathrm{min})$  for events with multiplicity greater than a given value. The $S_\mathrm{T}$ variable \cite{exo-12-009} is defined as the sum of the magnitudes of the transverse momenta $p_\mathrm{T}$ of all objects satisfying the kinematic cuts in Table \ref{tab-kcuta} plus the missing energy $\slashed E_\mathrm{T}>50$ GeV, where $\slashed E_\mathrm{T}$ is the magnitude of the vector sum of the $p_\mathrm{T}$ of all objects. The event multiplicity $N$ is defined as the number of final-state objects used to calculate $S_\mathrm{T}$. The distributions of the partial cross section $\sigma(S_\mathrm{T}>S_\mathrm{T}^\mathrm{min})$ times acceptance $A$ (=100\%) \cite{Hou:2015gba} are shown in Fig. \ref{fig-xsecAcca}.
They were obtained by running CATFISH with $n =4$ and $\verb+NP+=4$. The upper two panels in Fig. \ref{fig-xsecAcca} show  $\sigma(S_\mathrm{T}>S_\mathrm{T}^\mathrm{min})\times A$ as a function of $M_*$ for $\verb+XMIN+=\verb+QMIN+=3$ and $\alpha=0.7$ (left panel) or $\alpha=0.9$ (right panel). The lower panels show $\sigma(S_\mathrm{T}>S_\mathrm{T}^\mathrm{min})\times A$ as a function of \verb+XMIN+ at $\alpha=0.7$ (left panel) or $\alpha=0.9$ (right panel) and $M_*= 1.51$ TeV, corresponding to $M_D=3$ TeV for which the CMS Collaboration has determined the lower bound on BH mass. These results can be used to estimate the lower bounds on $M_*$ and \texttt{XMIN} by requiring that the simulated partial cross sections are smaller than the CMS upper cross section limits (dot-dashed curves). The top (bottom) panels of Fig. \ref{fig-xsecAcca} implies $M_*\gtrsim 1.9\, (0.8)$ TeV and $\texttt{XMIN}\gtrsim 3.8\, (1.6)$ for $\alpha = 0.7\, (0.9)$, respectively.

\begin{figure}[!h]

\center

\includegraphics[height=7.2cm]{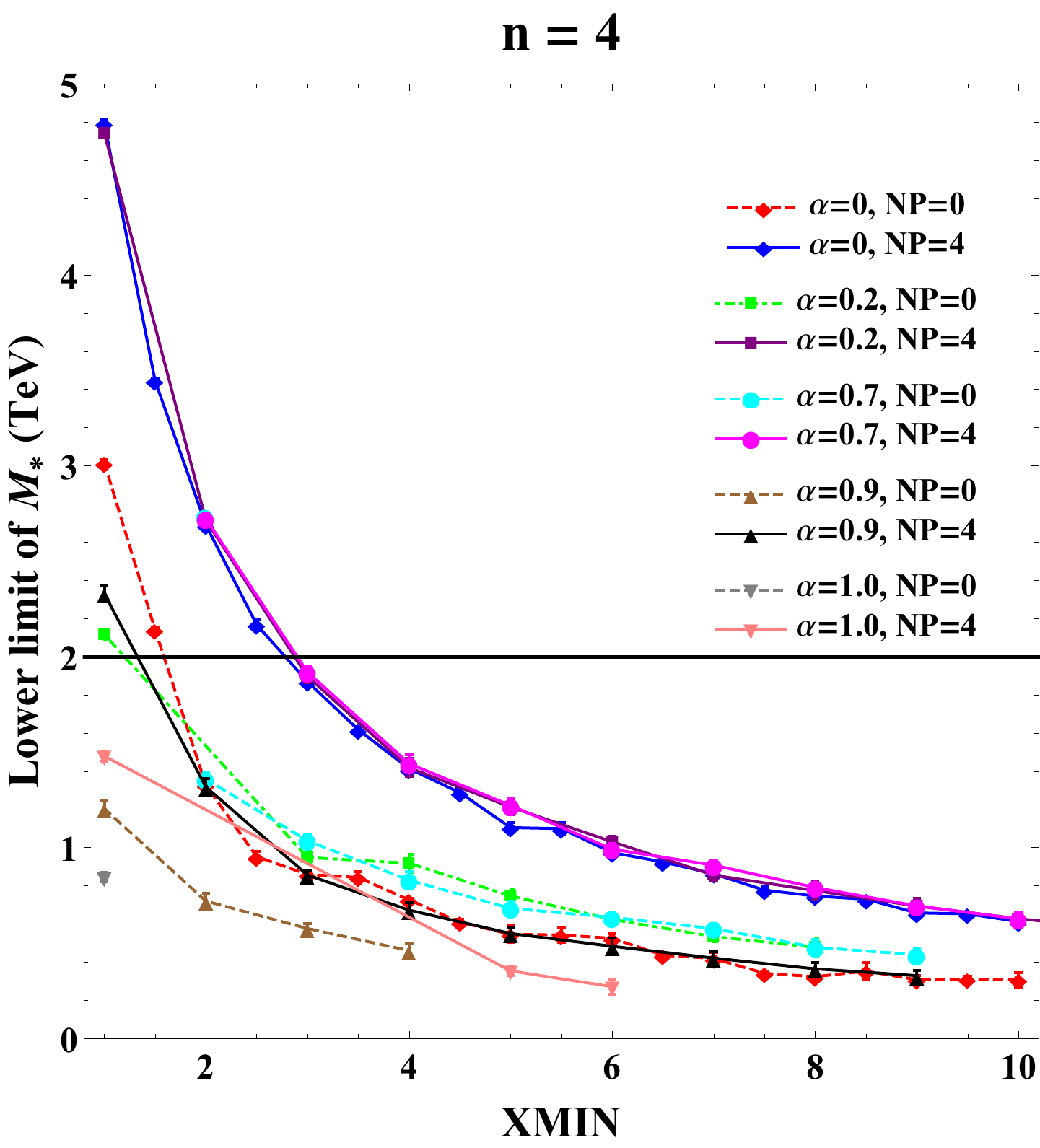}
\includegraphics[height=7.2cm]{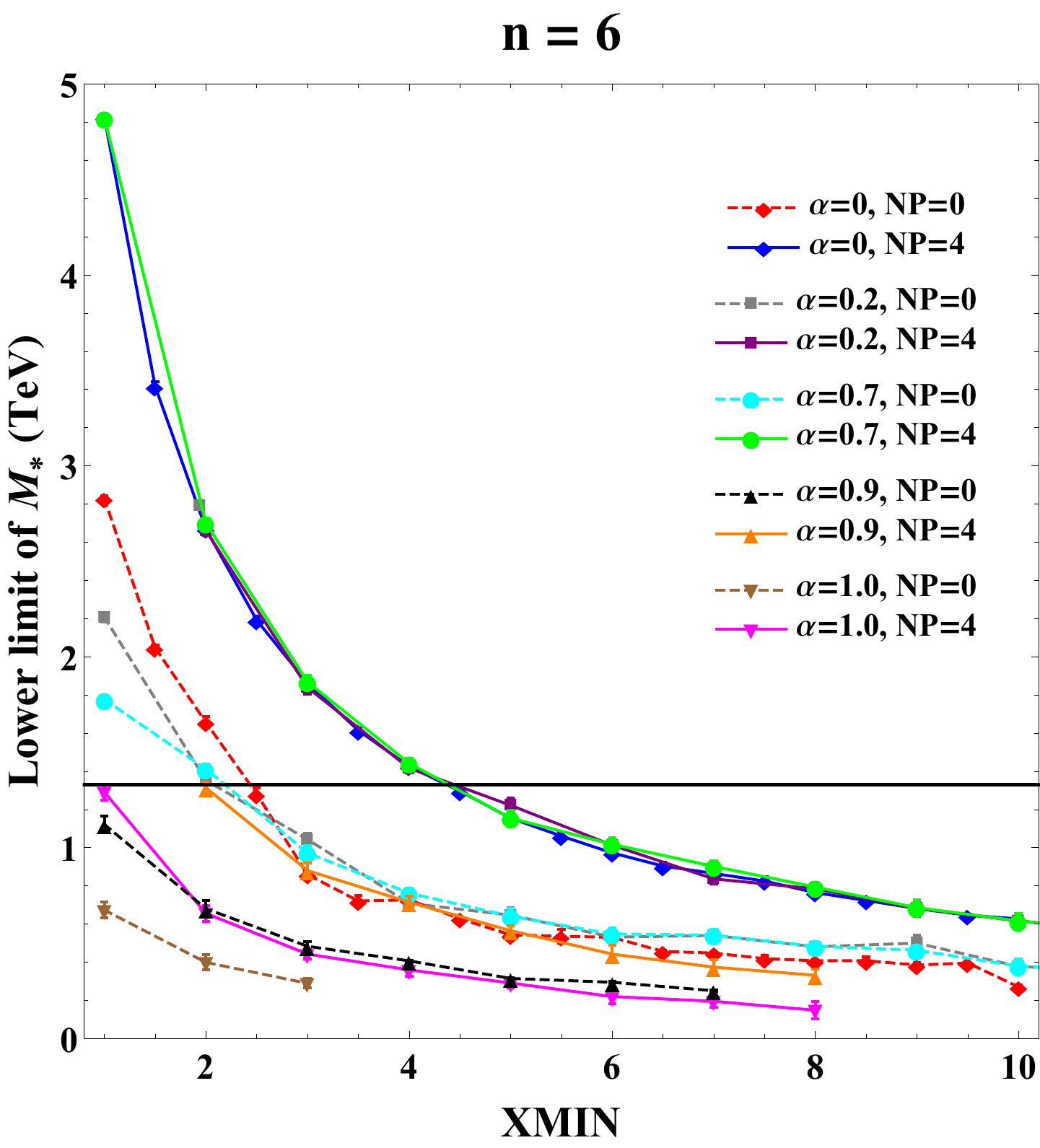}

\caption{\small Lower limit on $M_*$  as a function of \texttt{XMIN}, $\alpha$, \texttt{NP} and \texttt{NEXTRADIM} = 4 (left), and 6 (right). The dashed curves represent the results
for $\texttt{NP} = 0$ and the solid curves for $\texttt{NP} = 4$. The horizonal black lines represent the experimental limits from Ref.\cite{cmsepjc75}.}\label{amslmts}

\end{figure}

Figure \ref{amslmts} shows the lower bounds on $M_*$ as a function of \verb+XMIN+, \verb+NP+ and $\alpha$ for $n=4$ (left panel) and $n=6$ (right panel). Dashed lines are for $\verb+NP+=0$, and solid lines for $\verb+NP+=4$. The results for $\alpha=0$ give the bounds on $M_*$ in the absence of a minimum length \cite{Hou:2015gba}. As long as $\alpha\lesssim 0.7$, the bounds are not significantly different from the bounds when $\alpha=0$. However, when $\alpha \gtrsim 0.9$ the bounds on $M_*$ decrease rapidly as $\alpha$ increases with the cross section becoming highly suppressed when $M_\mathrm{ml}/M_*>1$. As in the standard scenario, $M_*$ bounds decrease as \texttt{XMIN} increases. The BH remnant model ($\texttt{NP}=0$) allows the production of BHs with smaller masses than models with $n$-body final decay ($\texttt{NP}\ne0$).

The results in Fig. \ref{amslmts} can be combined with lower bounds from other experiments to constrain
the minimum-allowed BH mass. The most stringent limits on the fundamental Planck scale $M_*$ from the analysis of monojet events are about 2 TeV and 1.33 TeV for $n=4$ and $n=6$, respectively \cite{cmsepjc75}. These bounds are represented by the horizontal lines in Fig. \ref{amslmts}. The left panel shows that for models with $\texttt{NP}=4$, $\texttt{XMIN}\lesssim3$ for $\alpha\le0.7$, while $\texttt{XMIN}\lesssim1$ for $\alpha=0.9$. At the same time, for models with $\texttt{NP}=0$, the upper bound of $\texttt{XMIN}$ is about 1-2 for $\alpha\le0.7$. Similarly, the right panel shows that for models with $\texttt{NP}=4$, $\texttt{XMIN}\lesssim4.5$ for $\alpha\le0.7$, while $\texttt{XMIN}\lesssim2$ for $\alpha=0.9$. For models with $\texttt{NP}=0$, the upper bound of $\texttt{XMIN}$ is about 2-2.5 for $\alpha\le0.7$. Note that these upper bounds on \texttt{XMIN} were obtained for the values actually explored in the simulation.

\begin{figure}[!h]
\center

\includegraphics[height=7.2cm]{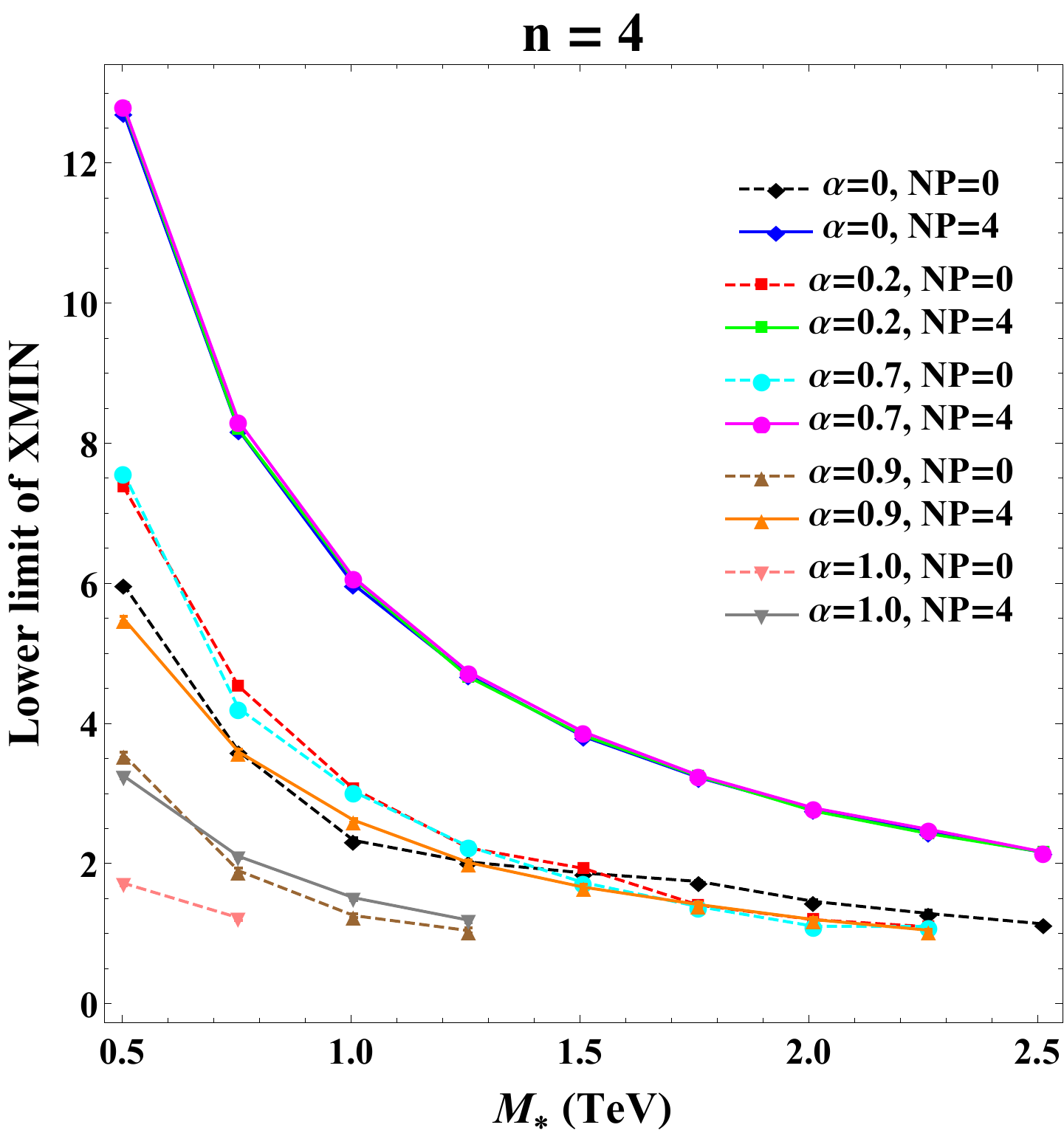}
\includegraphics[height=7.2cm]{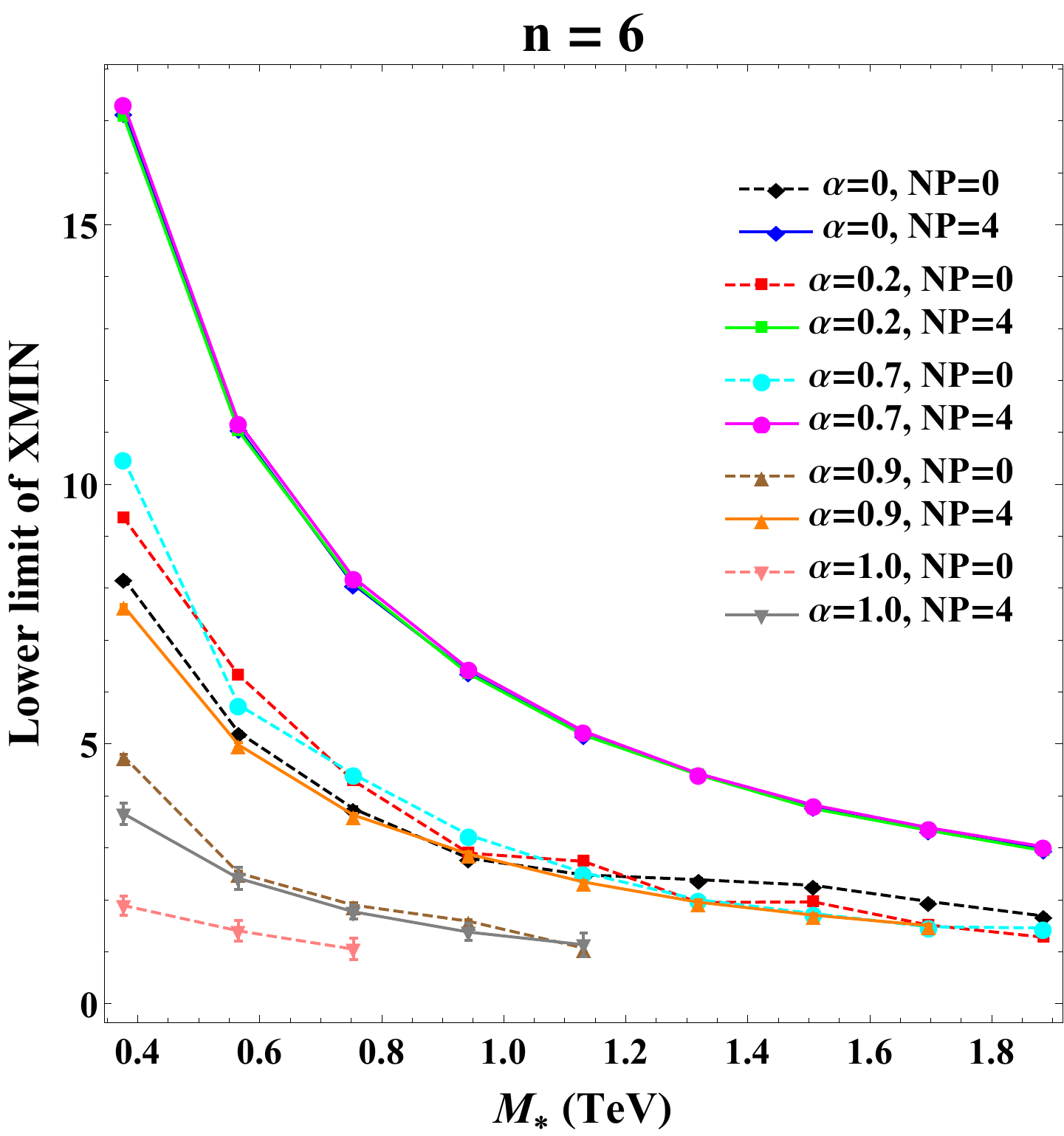}

\caption{\small SLower limit on \texttt{XMIN}  as a function of $M_*$, $\alpha$, \texttt{NP} and \texttt{NEXTRADIM} = 4 (right), and 6 (left). The dashed curves represent the results for $\texttt{NP} = 0$ and the solid curves for $\texttt{NP} = 4$.}\label{axminlmts}

\end{figure}

\begin{figure}[!ht]
\center
\includegraphics[height=7.2cm]{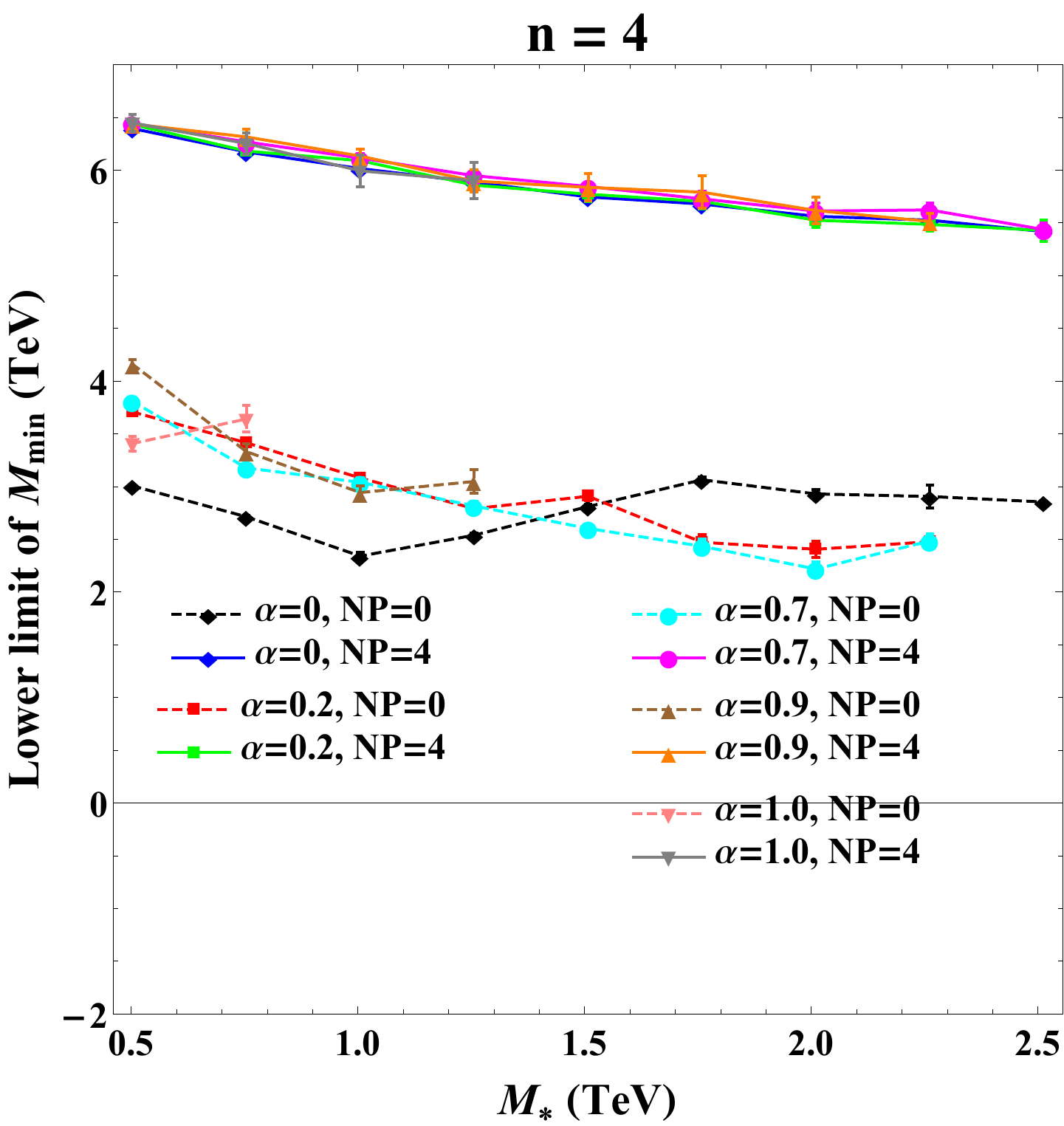}
\includegraphics[height=7.2cm]{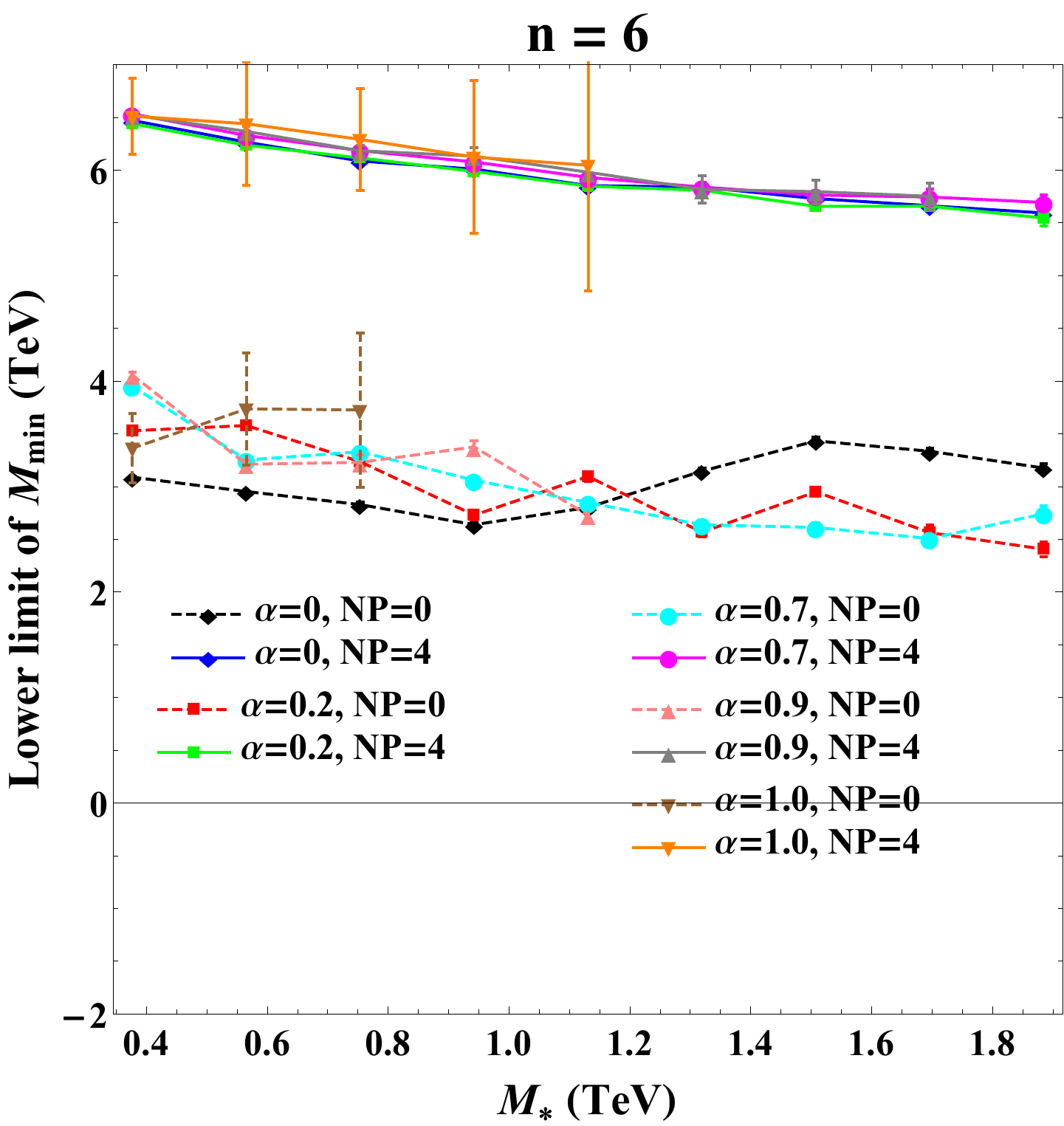}
\caption{\small Lower limit on $M_\mathrm{min}$  as a function of $M_*$, $\alpha$, \texttt{NP} and \texttt{NEXTRADIM} = 4 (right), and 6 (left).}
\label{fig-ambhlmts}

\end{figure}

Figure \ref{axminlmts} shows the lower bounds on \verb+XMIN+ as a function of $M_*$, \verb+NP+ and $\alpha$ for $n=4$ (left panel) and $n=6$ (right panel). When $\alpha\lesssim0.7$, there is no significant difference from $\alpha=0$. However, when $\alpha\gtrsim 0.9$, the bounds become smaller, due to a highly suppressed cross section. The BH remnant models (\texttt{NP}$=0$) give smaller lower bounds than the ``explosive" models (\texttt{NP}$\ne0$).
The lower bounds on \verb+XMIN+ decrease with $\alpha$ when $\alpha\gtrsim 0.9$. The lower bounds on $M_\mathrm{min}$ at $\verb+NP+=4$ do not depend significantly on $\alpha$'s (see Figure \ref{fig-ambhlmts}, because the cross section is determined by $M_\mathrm{min}$ as long as $M_*$ is fixed).

\begin{figure}[!h]

\center

\includegraphics[height=7.2cm]{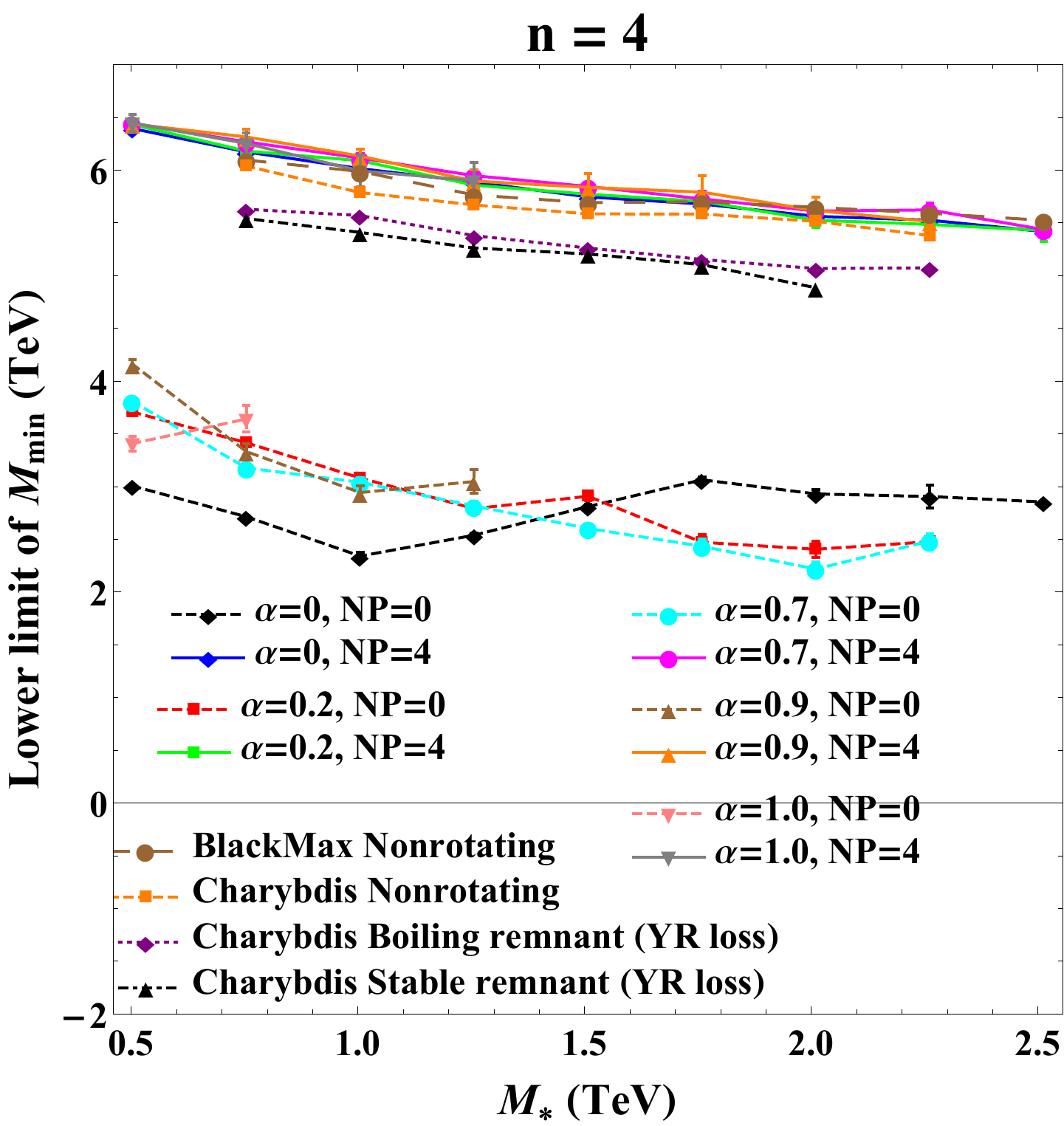}
\includegraphics[height=7.2cm]{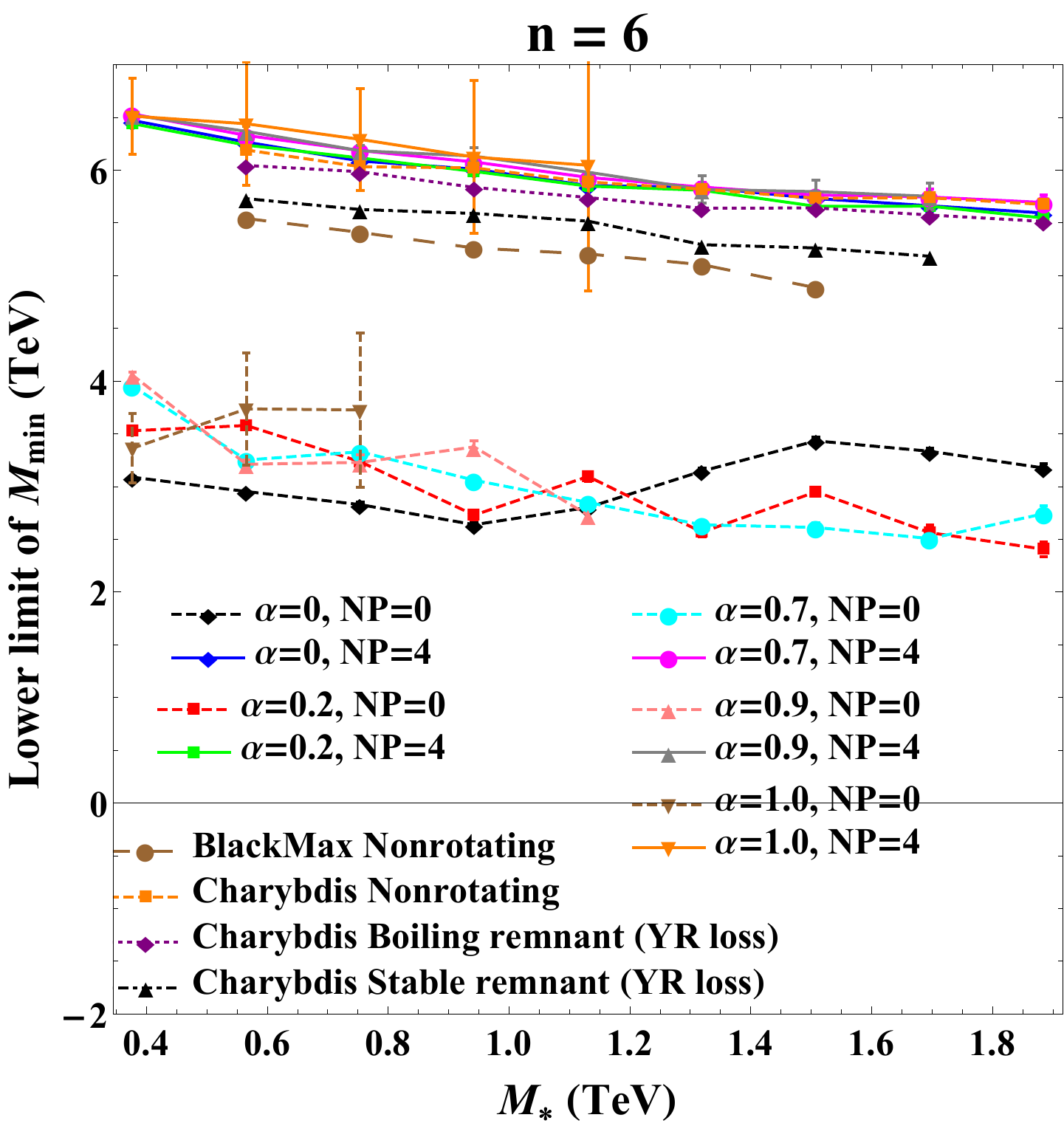}

\caption{\small Comparison of the lower bounds of $M_\mathrm{min}$ from CATFISH, BlackMax and CHARYBDIS2 for $n=4$ (left) and 6 (right). The results of BlackMax and CHARYBDIS2 are extracted from Figure 4 in Ref.\cite{exo-12-009}. }\label{acom}

\end{figure}

\begin{figure}[!h]

\center

\includegraphics[height=7.2cm]{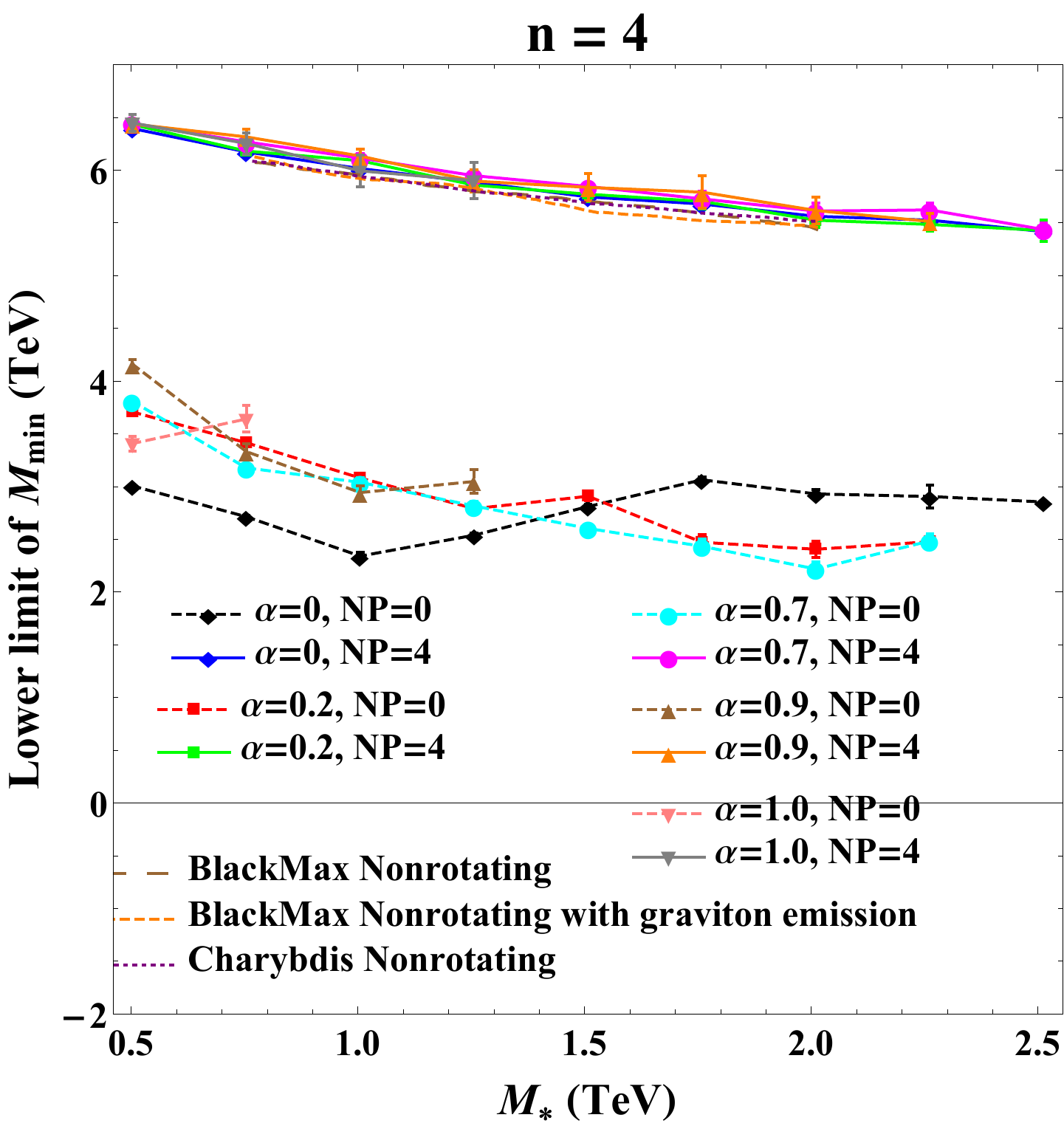}
\includegraphics[height=7.2cm]{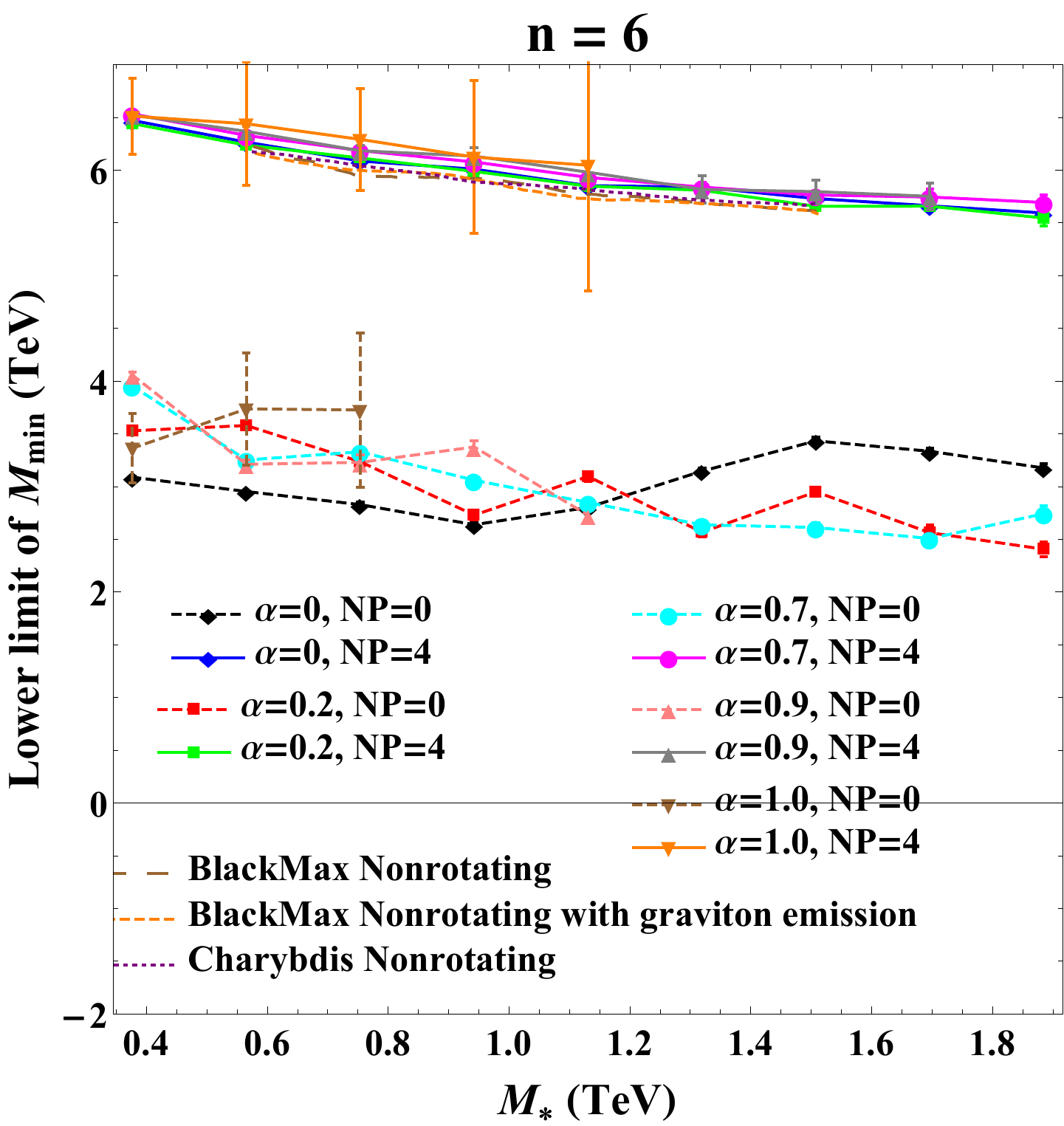}

\caption{\small Lower bounds on $M_\mathrm{min}$ from CATFISH, BlackMax and CHARYBDIS2 for $n=4$ (left) and 6 (right). The results of BlackMax and CHARYBDIS2 are extracted from Figure 's 8 and 10 in Ref.\cite{atlasjhep}. }\label{acom2}

\end{figure}

Figures \ref{acom}, \ref{acom2} show a comparison between the lower bounds on $M_\mathrm{min}$ from CATFISH with those from BlackMax \cite{Dai:2007ki,Dai:2009by} and CHARYBDIS2 \cite{Harris:2003db,Frost:2009cf}. The lower bounds  on $M_\mathrm{min}$ do not depend strongly on $\alpha$. Those for the remnant models (\texttt{NP}$=0$) are smaller than those predicted by BlackMax and CHARYBDIS2. The CATFISH ``explosive" models (\texttt{NP}$\ne0$) predict results very similar to those of the other two generators, except that the boiling/stable remnant models of CHARBDIS2 give slightly smaller limits than CATFISH, but still larger than the CATFISH remnant models (see Figure \ref{acom}). The difference in the predictions between CATFISH and CHARYBDIS2 with a stable remnant is due to the different implementations of the quantum phase by the two generators \cite{exo-12-009,Harris:2003db}. Moreover, CATFISH's stable remnant is invisible to the detector and contributes to missing energy, while CHARYBDIS2's remnant behaves as a heavy fundamental particle with conventional interactions in the detector. The three generators differ from one another also in the implementation of the quantum phase and in the inclusion or exclusion of the effect of gravitational energy loss at the formation of the BHs. For example, the predictions of the behavior of $M_\mathrm{min}$ vs. $M_*$ for \verb+NP+ = 0 (BH remnant) by CATFISH differ from those of BlackMax and CHARYBDIS2. The similarities among the three generators  are due to the fact that the three generators incorporate the same basic physics of microscopic BH formation and decay (see Fig. \ref{acom2}).

\section{Conclusion}\label{sec-final}

In this work, the effects of the GUP on BH production and decay at the LHC were investigated by simulating events with the BH event generator CATFISH. Upper BH production cross section limits are used to set lower bounds on the fundamental Planck scale $M_*$ and the minimum BH mass $M_\mathrm{min}$. The GUP decreases the production cross section as long as $\alpha$ is greater than a critical value $\alpha_c$, depending on $n$. The lower bounds on $M_*$ do not significantly change with respect to the standard scenario as long as $\alpha<\alpha_c$. A similar behavior is observed for the lower bounds on \texttt{XMIN}. Lower bounds on $M_\mathrm{min}$ are essentially unaffected by the GUP. For models without a BH stable remnant ($\texttt{NP}=4$) they generally agree with earlier bounds from the CMS Collaboration and the ATLAS Collaboration obtained with different BH event generators, i.e., BlackMax \cite{Dai:2007ki,Dai:2009by}, and CHARYDIS2 \cite{Harris:2003db,Frost:2009cf}. BH remnant models give milder constraints than non-remnant models. In summary, the GUP gives more stringent constraints on the sizes of the LEDs ($\lesssim 1.10$ nm), for $\alpha>\alpha_c$.

\section{Acknowledgements}

We wish to thank UAHPC \cite{rc2} at the University of Alabama and the Alabama Supercomputer Authority \cite{asc} for providing the computing infrastructure essential to our analysis.


\newpage

\begin{thebibliography}{10}

\bibitem{aad98}
I.~Antoniadis, N.~Arkani-Hamed, S.~Dimopoulos, and G.~R. Dvali.
\newblock {New dimensions at a millimeter to a Fermi and superstrings at a
  TeV}.
\newblock {\em Phys. Lett.}, B436:257--263, 1998.

\bibitem{add98}
N.~Arkani-Hamed, S.~Dimopoulos, and G.~R. Dvali.
\newblock {The Hierarchy problem and new dimensions at a millimeter}.
\newblock {\em Phys. Lett.}, B429:263--272, 1998.

\bibitem{add99}
N.~Arkani-Hamed, S.~Dimopoulos, and G.~R. Dvali.
\newblock {Phenomenology, astrophysics and cosmology of theories with
  submillimeter dimensions and TeV scale quantum gravity}.
\newblock {\em Phys. Rev.}, D59:086004, 1999.

\bibitem{AMATI198941}
D.~Amati, M.~Ciafaloni, and G.~Veneziano.
\newblock Can spacetime be probed below the string size?
\newblock {\em Physics Letters B}, 216(1):41 -- 47, 1989.

\bibitem{AMATI1990550}
D.~Amati, M.~Ciafaloni, and G.~Veneziano.
\newblock Higher-order gravitational deflection and soft bremsstrahlung in
  planckian energy superstring collisions.
\newblock {\em Nuclear Physics B}, 347(3):550 -- 580, 1990.

\bibitem{AMATI1993707}
D.~Amati, M.~Ciafaloni, and G.~Veneziano.
\newblock Effective action and all-order gravitational eikonal at planckian
  energies.
\newblock {\em Nuclear Physics B}, 403(3):707 -- 724, 1993.

\bibitem{Konishi1990276}
K.~Konishi, G.~Paffuti, and P.~Provero.
\newblock Minimum physical length and the generalized uncertainty principle in
  string theory.
\newblock {\em Physics Letters B}, 234(3):276 -- 284, 1990.

\bibitem{Maggiore:1993zu}
M.~Maggiore.
\newblock {Quantum groups, gravity and the generalized uncertainty principle}.
\newblock {\em Phys. Rev.}, D49:5182--5187, 1994.

\bibitem{Maggiore:1993kv}
M.~Maggiore.
\newblock {The Algebraic structure of the generalized uncertainty principle}.
\newblock {\em Phys. Lett.}, B319:83--86, 1993.

\bibitem{Rovelli:1994ge}
C.~Rovelli and L.~Smolin.
\newblock {Discreteness of area and volume in quantum gravity}.
\newblock {\em Nucl. Phys.}, B442:593--622, 1995.
\newblock [Erratum: Nucl. Phys.B456,753(1995)].

\bibitem{hawkeff}
S.~W. Hawking.
\newblock Particle creation by black holes.
\newblock {\em Comm. Math. Phys.}, 43(3):199--220, 1975.

\bibitem{kt}
K.~S. Thorne.
\newblock {\em {Magic Without Magic: John Archibald Wheeler}}.
\newblock Freeman, San Francisco, 1972.

\bibitem{mpDbh}
R.~C. {Myers} and M.~J. {Perry}.
\newblock {Black holes in higher dimensional space-times}.
\newblock {\em Annals of Physics}, 172:304--347, December 1986.

\bibitem{mc}
M.~Cavagli\`a.
\newblock {Black hole and brane production in TeV gravity: A Review}.
\newblock {\em Int. J. Mod. Phys.}, A18:1843--1882, 2003.

\bibitem{kantiReview}
P.~Kanti.
\newblock {Black holes in theories with large extra dimensions: A Review}.
\newblock {\em Int. J. Mod. Phys.}, A19:4899--4951, 2004.

\bibitem{giddings}
S.~B. Giddings and S.~D. Thomas.
\newblock {High-energy colliders as black hole factories: The End of short
  distance physics}.
\newblock {\em Phys. Rev.}, D65:056010, 2002.

\bibitem{gl}
G.~L. Landsberg.
\newblock {Black Holes at Future Colliders and Beyond}.
\newblock {\em J. Phys.}, G32:R337--R365, 2006.

\bibitem{Khachatryan:2010wx}
V.~Khachatryan et~al.
\newblock {Search for Microscopic Black Hole Signatures at the Large Hadron
  Collider}.
\newblock {\em Phys. Lett.}, B697:434--453, 2011.

\bibitem{CMS:2012yf}
S.~Chatrchyan et~al.
\newblock {Search for narrow resonances and quantum black holes in inclusive
  and $b$-tagged dijet mass spectra from $pp$ collisions at $\sqrt{s}=7$ TeV}.
\newblock {\em JHEP}, 01:013, 2013.

\bibitem{Chatrchyan:2012taa}
S.~Chatrchyan et~al.
\newblock {Search for microscopic black holes in $pp$ collisions at
  $\sqrt{s}=7$ TeV}.
\newblock {\em JHEP}, 04:061, 2012.

\bibitem{exo-12-009}
S.~Chatrchyan et~al.
\newblock {Search for microscopic black holes in $pp$ collisions at $\sqrt s$ =
  8 TeV}.
\newblock {\em JHEP}, 07:178, 2013.

\bibitem{Khachatryan:2015sja}
V.~Khachatryan et~al.
\newblock {Search for resonances and quantum black holes using dijet mass
  spectra in proton-proton collisions at $\sqrt{s} =$ 8 TeV}.
\newblock {\em Phys. Rev.}, D91(5):052009, 2015.

\bibitem{catfish}
M.~Cavagli\`a, R.~Godang, L.~Cremaldi, and D.~Summers.
\newblock {Catfish: A Monte Carlo simulator for black holes at the LHC}.
\newblock {\em Comput. Phys. Commun.}, 177:506--517, 2007.

\bibitem{PhysRev.71.38}
H.~S. Snyder.
\newblock Quantized space-time.
\newblock {\em Phys. Rev.}, 71:38--41, Jan 1947.

\bibitem{PhysRev.135.B849}
C.~A. Mead.
\newblock Possible connection between gravitation and fundamental length.
\newblock {\em Phys. Rev.}, 135:B849--B862, Aug 1964.

\bibitem{PhysRev.143.990}
C.~A. Mead.
\newblock Observable consequences of fundamental-length hypotheses.
\newblock {\em Phys. Rev.}, 143:990--1005, Mar 1966.

\bibitem{MAJID1994348}
S.~Majid and H.~Ruegg.
\newblock Bicrossproduct structure of $\kappa$-poincare group and
  non-commutative geometry.
\newblock {\em Physics Letters B}, 334(3):348 -- 354, 1994.

\bibitem{Hossenfelder:2012jw}
S.~Hossenfelder.
\newblock {Minimal Length Scale Scenarios for Quantum Gravity}.
\newblock {\em Living Rev. Rel.}, 16:2, 2013.

\bibitem{Cavaglia:2003qk}
M.~Cavagli\`a, S.~Das, and R.~Maartens.
\newblock {Will we observe black holes at LHC?}
\newblock {\em Class. Quant. Grav.}, 20:L205--L212, 2003.

\bibitem{Cavaglia:2004jw}
M.~Cavagli\`a and S.~Das.
\newblock {How classical are TeV scale black holes?}
\newblock {\em Class. Quant. Grav.}, 21:4511--4522, 2004.

\bibitem{Sjostrand:2006za}
T.~Sjöstrand, S.~Mrenna, and P.~Z. Skands.
\newblock {PYTHIA 6.4 Physics and Manual}.
\newblock {\em JHEP}, 05:026, 2006.

\bibitem{Sjostrand:2014zea}
T.~Sjöstrand, S.~Ask, J.~R. Christiansen, R.~Corke, N.~Desai, P.~Ilten,
  S.~Mrenna, S.~Prestel, C.~O. Rasmussen, and P.~Z. Skands.
\newblock {An Introduction to PYTHIA 8.2}.
\newblock {\em Comput. Phys. Commun.}, 191:159--177, 2015.

\bibitem{Boos:2001cv}
E.~Boos et~al.
\newblock {Generic user process interface for event generators}.
\newblock In {\em {Physics at TeV colliders. Proceedings, Euro Summer School,
  Les Houches, France, May 21-June 1, 2001}}, 2001.

\bibitem{Alwall2007300}
J.~Alwall et~al.
\newblock A standard format for les houches event files.
\newblock {\em Computer Physics Communications}, 176(4):300 -- 304, 2007.

\bibitem{Hou:2015gba}
S.~Hou, B.~Harms, and M.~Cavagli\`a.
\newblock {Bounds on large extra dimensions from the simulation of black hole
  events at the LHC}.
\newblock {\em JHEP}, 11:185, 2015.

\bibitem{deFavereau:2013fsa}
J.~de~Favereau, C.~Delaere, P.~Demin, A.~Giammanco, V.~Lemaître, A.~Mertens,
  and M.~Selvaggi.
\newblock {DELPHES 3, A modular framework for fast simulation of a generic
  collider experiment}.
\newblock {\em JHEP}, 02:057, 2014.

\bibitem{Cacciari:2011ma}
M.~Cacciari, G.~P. Salam, and G.~Soyez.
\newblock {FastJet User Manual}.
\newblock {\em Eur. Phys. J.}, C72:1896, 2012.

\bibitem{cmsepjc75}
V.~Khachatryan et~al.
\newblock {Search for dark matter, extra dimensions, and unparticles in monojet
  events in proton–proton collisions at $\sqrt{s} = 8$ TeV}.
\newblock {\em Eur. Phys. J.}, C75(5):235, 2015.

\bibitem{Dai:2007ki}
D.-C. Dai, G.~Starkman, D.~Stojkovic, C.~Issever, E.~Rizvi, and J.~Tseng.
\newblock {BlackMax: A black-hole event generator with rotation, recoil, split
  branes, and brane tension}.
\newblock {\em Phys. Rev.}, D77:076007, 2008.

\bibitem{Dai:2009by}
D.-C. Dai, C.~Issever, E.~Rizvi, G.~Starkman, D.~Stojkovic, and J.~Tseng.
\newblock {Manual of BlackMax, a black-hole event generator with rotation,
  recoil, split branes, and brane tension}.
\newblock 2009.

\bibitem{Harris:2003db}
C.~M. Harris, P.~Richardson, and B.~R. Webber.
\newblock {CHARYBDIS: A Black hole event generator}.
\newblock {\em JHEP}, 08:033, 2003.

\bibitem{Frost:2009cf}
J.~A. Frost, J.~R. Gaunt, M.~O.~P. Sampaio, M.~Casals, S.~R. Dolan, M.~A.
  Parker, and B.~R. Webber.
\newblock {Phenomenology of Production and Decay of Spinning Extra-Dimensional
  Black Holes at Hadron Colliders}.
\newblock {\em JHEP}, 10:014, 2009.

\bibitem{atlasjhep}
G.~Aad et~al.
\newblock {Search for microscopic black holes and string balls in final states
  with leptons and jets with the ATLAS detector at $\sqrt s$ = 8 TeV}.
\newblock {\em JHEP}, 08:103, 2014.

\bibitem{rc2}
University of~Alabama.
\newblock Research computing.
\newblock \url{https://researchcomputing.ua.edu/}, 2010.
\newblock Used since 2011.

\bibitem{asc}
Alabama supercomputer center (asc), 1989.

\end{thebibliography}
\end{document}